\documentclass[%
reprint,
superscriptaddress,
 amsmath,amssymb,
 aps,
prb,
]{revtex4-1}
\usepackage{physics}
\usepackage{float}
\usepackage{braket}
\usepackage{graphicx}
\usepackage{dcolumn}
\usepackage{bm}

\begin{document}
\preprint{APS/123-QED}
\title{Topological Corner States on Kagome Lattice Based Chiral Higher-Order Topological Insulator }

\author{Yichen Xu}
\affiliation{Department of Modern Physics, University of Science and Technology of China, Hefei, China}
\author{Ruolan Xue}
\affiliation{Department of Physics, University of Science and Technology of China, Hefei, China}
\author{Shaolong Wan}
\affiliation{Department of Modern Physics, University of Science and Technology of China, Hefei, China}

\date{\today}

\begin{abstract}
The higher-order topological insulator (HOTI) protected by spacial symmetry has been studied in-depth on models with square lattice. Our work, based on an alternative model on the breathing Kagome lattice, revealed that the different types of corners in the lattice could actually be conditionally gapless, or always gapped. Using the Wilson loop formalism, we argue that these corner states occur when the eigenvalues of the Wannier Hamiltonian cross through a certain reference point during the conceptual ``pumping" procedure. The results demonstrate the corner of the Kagome lattice based HOTI is a zero-dimensional analogue of the 1D chiral edge states on the boundary of a Chern insulator, but with a sensitive dependence on the shape of the corner. Our method of the pumping cylinder, which reveals the symmetry/gapless-ability correspondence, can be generalized into a general scheme in determining the classification of corner(hinge) states in HOTI.
\end{abstract}
\maketitle


\section{Introduction}
The symmetry protected first-order topological phases\cite{chiu2016classification,chen2013symmetry,chen2012symmetry,schnyder2008classification} have came into our sights for a long time. These phases are protected by intrinsic symmetries including time reversal, charge conjugation and chiral symmetry for the Altland-Zirnbauer(AZ) classes\cite{altland1997nonstandard} or more generalized on-site symmetries\cite{chen2013symmetry}. According to the Lieb-Schutz-Mattis Theorem, the symmetry-preserving boundary states of these phases can only be gapless, or gapped with topological order under strong interaction\cite{lieb1961two,vishwanath2013physics,wang2014classification,oshikawa2000commensurability,hastings2004lieb}. Such symmetries for protection have been generalized into crystalline symmetries\cite{fu2011topological,slager2013space,isobe2015theory,qi2015anomalous,kruthoff2016topological,hong2017topological,song2017topological,cheng2017microscopic,lu2017classification}, including rotational symmetries or space-group symmetries, and time translation symmetries\cite{rechtsman2013photonic,lindner2011floquet,nathan2015topological,von2016phase,potter2016classification,else2016classification,yan2017floquet}. The existence of the gapless boundary states originates from the non-trivial topological numbers\cite{fu2007topological,fu2007topologicalinv,morimoto2013topological} carried by Bloch bands (or Floquet-Bloch bands) in the bulk, like $\mathbb{Z}_2$ Fu-Kane invariant for 3D topological insulator in AZ class AII or $\mathbb{Z}$ classification for 3D class AIII. These numbers induce charge polarizations which can be captured by the quantized topological $\theta$ term\cite{qi2011topological,ryu2012electromagnetic,PhysRevB.78.195424,xu2013nonperturbative}. All of these phases have been classified theoretically using Clifford algebra(K-theory)\cite{morimoto2013topological,kruthoff2016topological}, cohomological approach\cite{chen2013symmetry} or directly using the topological terms in the effective actions\cite{lu2012theory,bi2015classification}.

The recent proposal of the higher-order topological insulators(HOTI) and superconductors(HOTSc) has broadened our concept of symmetry protected topological phases\cite{sessi2016robust,benalcazar2017quantized,schindler2017higher,song2017d,langbehn2017reflection,benalcazar2017electric,imhof2017topolectrical,ezawa2017higher}. Rather than having gapless symmetry-preserving boundary states, these phases have gapped edge states at the (d-1)D boundary of a d dimensional bulk with non-trivial topology, and these gapped boundary states possess gapless while symmetry-preserving hinge states at the (d-2)D boundary of the (d-1)D boundary. This hierarchical topological behaviors are induced by nonzero polarization qaudrupoles in 2D or octupoles in 3D. The existence of these higher-order polarization is protected by compositional anti-unitary symmetry groups, which endows the two edges of the hinge with different non-trivial mirror Chern numbers, causing the hinge to be gapless. To understand their origin, we can start from a first order TI with gapless boundary states protected by inversion symmetries, and adding symmetry breaking gapping terms to the effective boundary Dirac theories. These terms are usually odd under reflection\cite{langbehn2017reflection}, causing the effective mass of the two boundaries of a hinge having opposite signs, thus the flip of the mass gives rise to gapless hinge states. Viewing from the bulk, the terms added to gap the boundary break certain rotation symmetries($C_4$ for previous works), which can be restored by making these symmetry operations anti-unitary, like $C_4\mathcal{T}$.

In this paper we consider a new proposal of HOTI on breathing Kagome lattice\cite{ezawa2017higher}. The model posses three types of inequivalent corners, rather than only one (the right angle $\pi/2$) for square lattice. These three types of corners are calculated to be either conditionally gapless or always gapped. This unusual dependence of the gapless-ability on the shape of the corner, which previous works do not capture, goes beyond previous understandings\cite{langbehn2017reflection,schindler2017higher}. We provide a explicit proof of this correspondence relationship between gapless-ability and the shape using the Wilson loops under symmetry transformations. This proof is also a classification scheme of a lattice system with any point group symmetries in two spacial dimensions. The paper is organized as follows: In Sec. II we summarize the symmetries of the HOTI model on breathing Kagome lattice, and clarify the concept of a ``boundary" on such lattice. In Sec. III we performed numerical calculations of the model in rhombus, hexagonal and ribbon geometry. The numerical diagonalization of the first two geometries reveals the crucial shape dependence we will discuss in the rest of the paper, and the band calculation in ribbon geometry will elicit an apparent paradox which can only be solved by using Wilson loop formalism. In Sec. IV we introduce the Wilson loop in the model, and give the argument to the symmetry/gapless-ability correspondence, which resolves the paradox in Sec. III at the same time. The two appendices are complementary details and proofs we need for argument in Sec. IV, and the window towards a general classification scheme.
    \begin{figure}[H]
	\includegraphics[width=7.5cm]{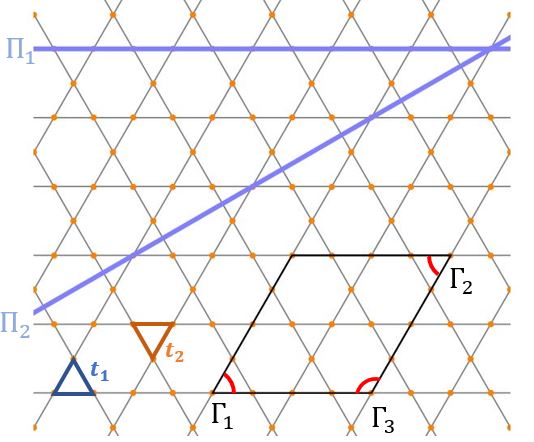}
	\caption{The figure shows upward (blue) and downward (orange) triangles, two safe choices of edges and three types of corners. A rhombus in the lattice always has one $\Gamma_1$, one $\Gamma_2$ and two $\Gamma_3$ corners, as can be seen from the figure.}
\end{figure}
\section{The model, symemtries, corners and edges}
The Ezawa's HOTI model\cite{ezawa2017higher} on breathing Kagome lattice can be written as 
\begin{equation}
    H=-
\sum_{\langle ij\rangle}t_p c_i^\dagger c_j
\end{equation}
where the hopping strengths $t_p=t_{1(2)}$ if the bond between $i$ and $j$ lives on the triangle pointing upward(downward). (See Fig. 1.) The bulk tight-binding Hamiltonian can be written as 
\begin{equation}
    H(\mathbf{k})=-\sum_{i=1}^{8}\alpha_{i}\Lambda_{i}
\end{equation}
where $\{\Lambda_i\}$ are the eight Gell-Mann matrices and 
  $ \alpha_{1}=(t_1+t_2)\cos(k_x/2+\sqrt{3}k_y/2),
   \alpha_{2}=(-t_1+t_2)\sin(k_x/2+\sqrt{3}k_y/2),
   \alpha_{4}=(t_1+t_2)\cos k_x,
   \alpha_{5}=(-t_1+t_2)\sin k_x,
   \alpha_{6}=(t_1+t_2)\cos(k_x/2-\sqrt{3}k_y/2),
   \alpha_{7}=(-t_1+t_2)\sin(k_x/2-\sqrt{3}k_y/2)$ and $
   \alpha_{3,8}=0$. The model has simple form of time-reversal symmetry $\mathcal{T}H(\mathbf{k})\mathcal{T}^{-1}=H(-\mathbf{k})$ with $\mathcal{T}=\mathcal{K}$, along with the $\mathcal{C}_3$ rotation symmetry $\mathcal{C}_3 H(\mathbf{k})\mathcal{C}_3^{-1}=H(\mathcal{R}_3\mathbf{k})$ with $\mathcal{C}_3=\begin{pmatrix}
   0&1&0\\
   0&0&1\\
   1&0&0\\
   \end{pmatrix}$, where $\mathcal{R}_3$ stands for anti-clockwisely rotating $\mathbf{k}$ by $2\pi/3$. These two symmetries apparently commute. In contrast, the action $\mathcal{C}_6$ is NOT a symmetry for the $t_1\neq t_2$ case, which can be directly observed from the lattice structure of Kagome lattice, in the sense that if we rotate the lattice by $\pi/3$, the triangle pointing upward will point downward. However, it can be shown that the composition $\tilde{\mathcal{C}}_6=\mathcal{C}_3^{-1}\mathcal{T}$ is actually the symmetry of the Hamiltonian with the momentum $\mathbf{k}$ rotated by $\pi/3$. (We use a tilde to indicate the action is actually anti-unitary.) This is quite similar to the previously proposed model on square lattice (where $\mathcal{T}$ and $\mathcal{C}_4$ symmetry is broken but $\mathcal{C}_4\mathcal{T}$ is preserved). Below we shall see this compositional symmetry is crucial for the existence of non-trivial corner state.  Additionally, the system also has a mirror symmetry $\mathcal{M}_y H(\mathbf{k})\mathcal{M}_y^{-1}=H(\mathcal{R}_y k)$,where $\mathcal{R}_y(k_x,k_y)=(-k_x,k_y)$ and $\mathcal{M}_y=\begin{pmatrix}
   0&0&1\\
   0&1&0\\
   1&0&0\\
   \end{pmatrix}$ for any $t_1$ and $t_2$. Similar to the $C_3$ case, the mirror symmetry for $x$-axis is also compositional, which reads $\tilde{\mathcal{M}}_x=\mathcal{M}_y\mathcal{T}$. The symmetries satisfy $\tilde{\mathcal{C}}_6^2=\mathcal{C}_3$ and $\tilde{\mathcal{M}}_x^2=1$.

   From Fig. 1 we can see there are three different types of corners, denoted by $\Gamma_{1,2,3}$. $\Gamma_{1(2)} $ is the corner of the upward(downward) triangle, and $\Gamma_3$ is the corner of a hexagon. The existence of non-trivial corner state of the model gives us an important caveat that we should be careful about the definition of an ``edge" or ``boundary", because the corner at the edge/boundary can be regarded as the boundary of the edge/boundary, which may host nontrivial corner states. As we can see from Sec. III, the corner $\Gamma_3$ will not host any gapless corner state no matter what $t_1$ and $t_2$ are. Thus, the only two safe choices of boundary among all the possible edges on the Kagome lattice are the two indicated in Fig. 1, and we denote them by $\Pi_{1}$ and $\Pi_{2}$. Also note the reflection about $\Pi_2$ is unitary, because the directions of the triangles are not flipped as the case for $\Pi_1$.

   For edge $\Pi_1$, there is a distinction between whether the bulk is located at the upper or lower side of the edge, because clearly if we choose the upper side as the bulk, then the row of triangles adjacent to the edge on the other side of the bulk have hopping strength $t_2$, but the other case these adjacent triangles have hopping strength $t_1$. This originates from the breaking of the reflection symmetry $\mathcal{M}_x$, or so called ``breathing anisotropy"\cite{schaffer2017quantum}. Another feature worth notice is, this model does not have a first-order TI limit when the symmetry breaking terms are removed. In our case, such condition is recaptured at $t_1=t_2$, under which the bulk becomes gapless.
      \begin{figure}
   	\includegraphics[width=8.7cm]{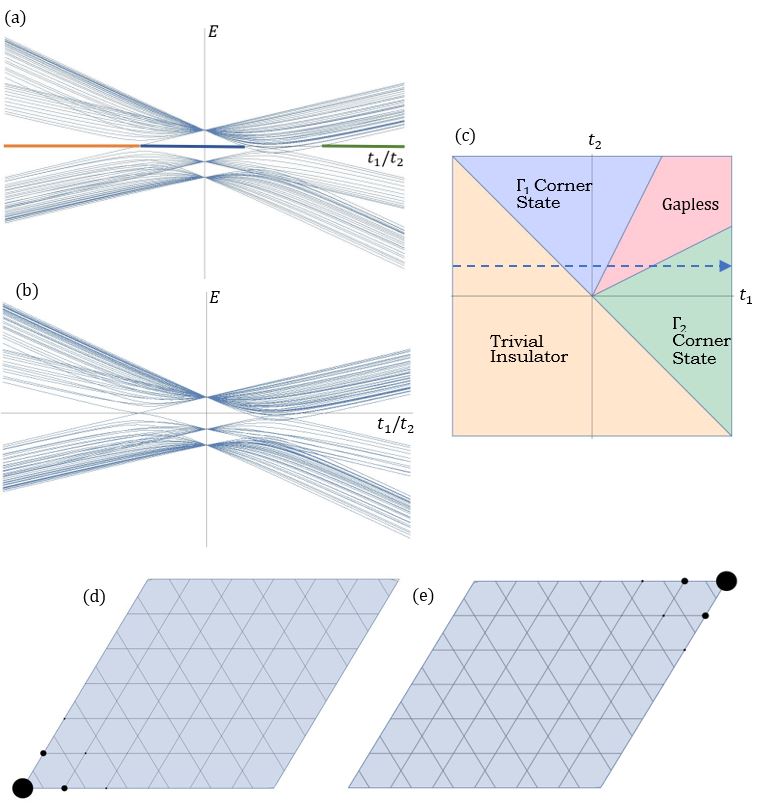}
   	\caption{(a) The energy spectrum of a rhombus with edge length $12a$ plotted as a function of $t_1/t_2$. There is always a zero energy state at every $t_1/t_2$. The cases of trivial insulator, gapless $\Gamma_1$, gapless $\Gamma_2$ region are indicated in orange, blue and green, correspondingly. (b) The energy spectrum of a hexagon obtained by removing two atoms at the $\pi/3$ angles. There is no mid-gap state if the spectrum is gapped (i.e. $t_1\neq- t_2$ and $t_1<t_2/2$ or $t_1>2t_2$. (c) The phase diagram of a rhombus. The process of tuning  $t_1/t_2$ in (a), thus driving the rhombus across the four phases, can be recovered if we go along the arrow in the diagram. (d,e) The root-modular-square of the wave function on each site of the rhombus, the parameters are in (d)$t_1=0.3t_2$. and (e)$t_1=3t_2$.}
   \end{figure}
   
   \section{Numerical results on the boundary and corner states}
   To test the higher-order nature of the model, we performed numerical diagonalization with a rhombus geometry (shown in Fig. 1). The result (see Fig. 2a) shows there is always a zero energy state in the energy spectrum.  this zero energy state submerges into bulk energies when $t_1=t_2/2$, and move out from them after $t_1>2t_2$. From Fig. 2a it can also be seen that each state which crosses $E=0$ between $t_2/2<t_1<2t_2$ will cross it two times, one downward and one upward, so in the region $t_1>2t_2$, the zero energy state is still topologically non-trivial. We can also check the wave function of the zero energy state (see Fig. 2d and 2e), which indeed shows high concentration at the $\Gamma_1$ corner when $-t_2<t_1<t_2/2$, and at $\Gamma_2$ corner if $t_1>2t_2$.  The state crossing at $t_1=-t_2$ indicates the zero energy state in the region $t_1<-t_2$ is topologically trivial. As a comparison, we obtained the energy spectrum of a hexagon (see Fig. 2b), by removing the atoms at $\Gamma_1$ and $\Gamma_2$ corner. Thus all the six corners belong to the $\Gamma_3$ type. The spectrum lacks a zero energy state as in Fig 2a, which shows the corner $\Gamma_3$ is always gapped outside the gapless region $t_2/2<t_1<2t_2$. This means $\Gamma_3$ corner is just a turning of the boundary, with no nontrivial topological properties. The results of the numerical diagonalization can be summarized into a phase diagrams of the rhombus, see Fig. 2c. 
   

		\begin{figure}[H]
		\centering
		\includegraphics[width=8.5cm]{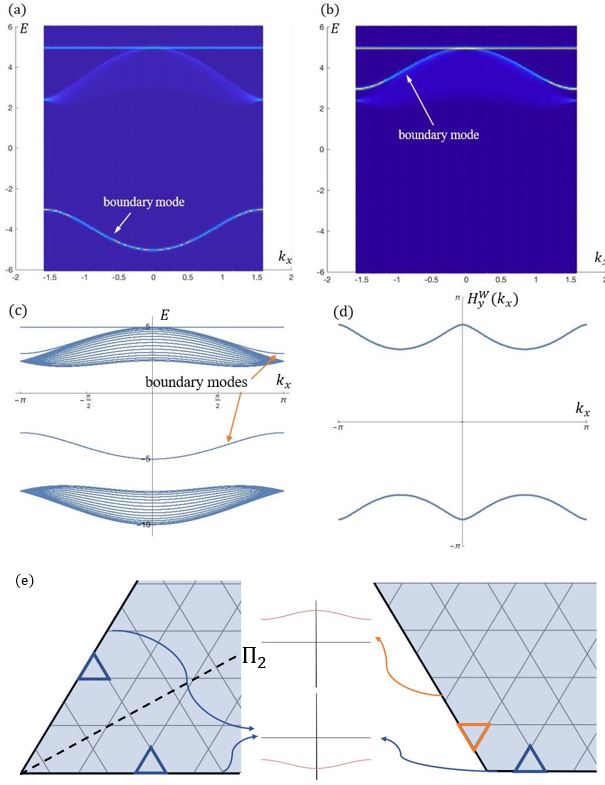}
		\caption{The bands of semi-infinite bulk calculated by tight-binding based iteration methods. The bulk is located on the upper side(a) and the lower side(b) of the boundary $\Pi_1$. The hopping parameters are chosen to be $t_1=1$ and $t_2=4$. The existence of boundary modes outside the bulk bands are indicated in the two cases. Note the signs of the boundary mode are opposite. This is a clear indication of the composite $\mathcal{M}_x\mathcal{T}$ symmetry. (c) The calculated band structure in ribbon geometry. (d) The eigenvalues of the Wannier Hamiltonian $H^W_y(k_x)$ as a function of $k_x$. The projection is onto the two non-flat bulk bands. (e) The paradox of having no band-crossing at $\Gamma_1$ corner. The bulk resides on the same side of the two edges of $\Gamma_1$ but different sides for $\Gamma_3$. The inset in the middle are sketches of boundary bands in (a) and (b).}
	\end{figure}
	We note that the corner can be regarded as the boundary of two non-parallel boundary lines. Thus, to give account to the existence of the gapless corner modes, we need to obtain the band structure at the presence of boundary. We conducted traditional calculation in ribbon geometry (Fig. 3c) and the tight-bonding based iteration techniques\cite{sancho1985highly,sancho1984quick} on the band in semi-infinite geometry, with the bulk resides on the upper and lower side of the boundary $\Pi_1$ (see Fig. 3a and 3b). In each case there exists one boundary mode outside the bulk spectrum, and the boundary modes coexists in the ribbon geometry. One reasonable conjecture here is the energy difference of boundary modes in the upper and lower bulk semi-infinite bands will lead to the gapless corner mode, like the situation in quantum Hall systems, when the crossing of the Landau level through the Fermi surface leads to the chiral boundary state. However, if we look carefully towards the lattice structure in Fig.1, we will find the two boundaries which cross each other at the corner $\Gamma_1$ and $\Gamma_2$ actually have same band structures (see Fig. 3e), because the bulk resides on the same side of boundary (with the triangles $t_1$ in the bulk adjacent to the two boundaries). Conversely, the two boundaries crossing at the corner $\Gamma_3$ have the bulk residing in the opposite direction. This means our band-crossing paradigm fails. Moreover, The ``mirror picture"\cite{schindler2017higher,langbehn2017reflection} of understanding the existence of gapless corner modes also fails here, because we can easily see the mirror plane of corners $\Gamma_{1,2}$ is $\Pi_2$, whose reflection symmetry does not break under the ``breathing anistropy", thus the boundary mass gap is indeed not flipped on the two edges. This paradox necessitates a deeper understanding of the existence of the boundary states.

	\section{The wilson-loop formalism and the symmetry arguments for the shape dependence}
	In first order TIs, the gapless nature of the boundary state can be easily understood through the ``band crossing" paradigm \cite{RevModPhys.83.1057,shen2012topological}, i.e. two bands, both exist in the TI bulk and vacuum, have opposite order in energy, thus cross at the boundary. This is induced by a non-trivial $\mathbb{Z}_2$ number of the TI bulk bands. In previous works\cite{benalcazar2017electric,schindler2017higher}, it is shown that the existence of the gapless hinge modes can also be understood by the ``band crossing" picture in a similar manner, in the sense that the boundaries on the two sides of the hinge have nontrivial ``mirror Chern numbers"\cite{morimoto2013topological}, causing the effective ``mass" on the two insulating boundaries have different signs. However, the boundary theories are captured in the Wilson loop picture, with the similar behavior happens in eigenspace of the so-called ``Wannier Hamiltonian". In this section we present the similar calculation of the boundary theory, but in aware of the different types of corners concerned (in contrast to the square lattice case with only one type of the $\pi/2$ corner).

	The Wilson loop matrix is defined to be
	\begin{equation}
	\label{wil}
	W_{mn}^y(k_x)=\bra{\psi_m(k_x,\frac{4\pi}{\sqrt{3}})}\prod_{ky=\frac{4\pi}{\sqrt{3}}}^{0}P(k_x,k_y)\ket{\psi_n(k_x,0)}
	\end{equation}
	where $n,m=1,2$ indicates the two non-flat bands of the bulk Hamiltonian and the projector
	\begin{equation}
	\label{proj}
	P(k_x,k_y)=\sum_{n=1}^{2}\ket{\psi_n(k_x,k_y)}\bra{\psi_n(k_x,k_y)}
	\end{equation}
	projects out the flat band. The loop wraps the first Brillouin zone in the $y$ direction(see Fig. 4). As predicted by \cite{yu2011equivalent}, the Wannier bands reproduce the properties of the gapped boundary modes in the ribbon geometry (see Fig. 3(d)).
	\begin{figure}[H]
		\centering
		\includegraphics[width=6cm]{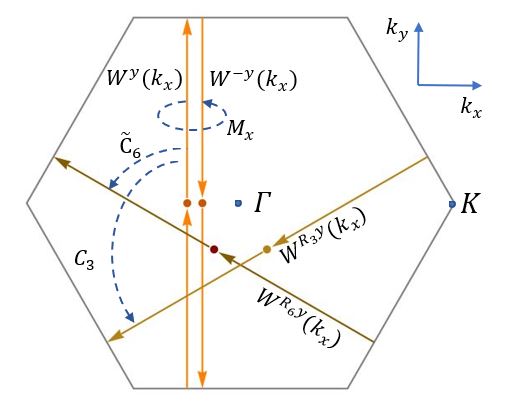}
		\caption{The Brillouin zone and the Wilson loops before and after symmetry transformations. The higher-symmetry points are $\Gamma=(0,0)$ and $K=(4\pi/3,0)$.}
	\end{figure}
	With the Wilson loop defined, we can now give account to the shape dependence of the corner states, which deeply rooted from the composite nature of the relevant symmetry groups mentioned in Sec. II. Follow the discussion in the appendices in \cite{benalcazar2017quantized}, we can obtain the action of symmetry transformations on the Wilson loop operator in Eq. (\ref{wil}).The eigenstates after the transformation $\tilde{\mathcal{M}}_x$ can be written as 
	\begin{equation}
	\begin{aligned}
	\tilde{\mathcal{M}}_x&\ket{\psi_n(k_x,k_y)}\\
	&=\ket{\psi_m(k_x,-k_y)}\bra{\psi_m(k_x,-k_y)}\tilde{\mathcal{M}}_x\ket{\psi_n(k_x,k_y)}\\
	&=\ket{\psi_m(k_x,-k_y)}M_{\mathbf{k},mn}^{x},
	\end{aligned}
	\end{equation}
	where $M_{\mathbf{k},mn}^{x}$ is the unitary sewing matrix 
	between original and transformed eigenstates. According to the orthogonal theorem, we can only sum the indices of the non-flat bands. Thus, the projector (\ref{proj}) is invariant under the reflecting action, which means under $\tilde{\mathcal{M}}$ we have $\tilde{\mathcal{M}}P(k_x,k_y)\tilde{\mathcal{M}}^{-1}=P(k_x,-k_y)$. Using this result we can write down how the Wilson loop matrix is transformed under the action $\tilde{\mathcal{M}}_x$, which reads

	\begin{equation}
	\begin{aligned}
	W^y_{mn}(k_x)=&M^{x\dagger}_{\mathbf{k},mk}W^{-y}_{kl}(k_x)M^{x}_{\mathbf{k},ln}\\
	=&M^{x\dagger}_{\mathbf{k},mk}W^{y\dagger}_{kl}(k_x)M^{x}_{\mathbf{k},ln}\\
	\end{aligned},
	\end{equation}
	or in a basis independent form
		\begin{equation}
	W^y(k_x)\to\tilde{\mathcal{M}}_xW^y(k_x)\tilde{\mathcal{M}}_x^{-1}=W^{-y}(k_x)=W^{y\dagger}(k_x).
	\end{equation}
	The part on the exponential $W^y(k_x)=e^{iH^y_W(k_x)}$ is usually called the Wannier Hamiltonian. Under reflection to the x axis, we have $\tilde{\mathcal{M}}_x H^y_W(k_x)\tilde{\mathcal{M}}_x^{-1}=H^{-y}_W(k_x)=-H^y_W(k_x)$. Clearly, this means the eigenvalues of the Wannier Hamiltonian comes in plus/minus pairs for each $k_x$, which can be easily seen from Fig. 3d. We denote the eigenstates of the Wilson loop operator (and thus the Wannier Hamiltonian) by $\ket{\varphi_\alpha(k_x)}(\alpha=1,2)$, and the corresponding eigenvalues of the Wannier Hamiltonian as $H^y_W(k_x)\ket{\varphi_\alpha(k_x)}=h^y_{W,\alpha}(k_x)\ket{\varphi_\alpha(k_x)}$, then under reflection we have
	\begin{equation}
	\begin{aligned}
	W^{-y}(k_x)\tilde{\mathcal{M}}_x\ket{\varphi_\alpha(k_x)}=&\tilde{\mathcal{M}}_x e^{ih^y_{W,\alpha}(k_x)}\ket{\varphi_\alpha(k_x)}\\
	=& e^{-ih^y_{W,\alpha}(k_x)}\tilde{\mathcal{M}}_x\ket{\varphi_\alpha(k_x)}
	\end{aligned}.
	\end{equation}
	The minus sign on the exponential is due to the anti-unitary nature of $\tilde{\mathcal{M}}_x$. Eq. (8) shows the eigenstates of the Wannier Hamiltonian are flipped under the $\tilde{\mathcal{M}}_x$. Using $W^{-y}(k_x)=W^{y\dagger}(k_x)=e^{-iH^y_W(k_x)}$ and comparing with Eq. (8), it can be seen that $\tilde{\mathcal{M}}_x\ket{\varphi_\alpha(k_x)}$ and $\ket{\varphi_\alpha(k_x)}$ actually have the same eigenvalue of $H^y_W(k_x)$.  This means the eigenvalues go back to their original place after we adiabatically pump up the $\tilde{\mathcal{M}}_x$ transformation, which implies the boundary is indeed gapped.
	
	From the above argument we can see that if the Hamiltonian do NOT go back to itself (like the case for $\tilde{\mathcal{M}}_x$), any anti-unitary operation will bring about the ``band crossing" we are pursuing, because the induced flip of the eigenstates, when the Wannier Hamiltonian is not flipped, gives birth to the crossing of eigenvalues through certain reference point, which can serve as the Fermi surface in the quantum Hall effect case. Below we shall see this is indeed the case for $\tilde{\mathcal{C}}_6$ symmetry.

	The similar relation to Eq. (7) is also valid for symmetries $\tilde{\mathcal{C}}_6$ and $\mathcal{C}_3$. We simply write them as
	\begin{equation}
	\begin{aligned}
	W^y(k_x)\to\tilde{\mathcal{C}}_6W^y(k_x)\tilde{\mathcal{C}}_6^{-1}=W^{R_6y}(R_6 k_x)\\
	W^y(k_x)\to\mathcal{C}_3 W^y(k_x)\mathcal{C}_3^{-1}=W^{R_3y}(R_3 k_x)\\
	\end{aligned}
	\end{equation}
	(see Fig. 4). Note the flip of the sign of the Wannier Hamiltonian also happens to $\tilde{\mathcal{C}}_6$ transformation. This may induce the ``band crossing" between the two hamiltonians $H^y_W(k_x)$ and $H^{R_6}_W(k_x)$ before and after transformation. In the new ``Wannier band insulator", because the Wannier bands actually live in $\mathbb{R}/(2\pi\mathbb{Z})$, if we choose a reference point in the ``Brillouin zone" $[-\pi,\pi)$, then the crossing of the Wannier bands before and after the symmetry transformation (which transforms one edge of the corner into the other) through the reference point will produce the 1D analog of the chiral edge mode which is just the single 0 eigenvalue in the spectrum. However, we should be extremely careful about how the bands are actually transformed under $\tilde{\mathcal{C}}_6$, because now we have additional possibilities other than a trivial cross. For the convenience of our future discussion, we define a ``pumping cylinder" here. We first pick up a representative $h_W^{y}(0)$ in the band to represent a quasi-flat ``Landau-level" in the spectrum. The cylinder then glues the bands before and after the certain transformation, like $\tilde{\mathcal{C}}_6$ for the current case. After specifying it's initial and final position, the representative, or the energy of an eigenstate, can move freely on the cylinder, mimicking a process which eigenvalues move under adiabatic pump. However, it should be pointed out that this so-called pumping is purely conceptual, because of the discrete nature of the transformation we want to capture. From the argument in Appendix. A, we know that under requirements of symmetries, there are ONLY two inequivalent crossing pictures on the pumping cylinder, showed in Fig. 5(a) and (b). The case for $\Gamma_1$ with $-t_2\leq t_1\leq t_2/2$, discussed in Sec. II is clearly the case (a), with single chiral zero mode is produced during the ``pumping". However, if we have crossings with different chiralities (the case shown in Fig. 5(b)), the net effect will be canceled, thus ruling out the possibility of having a gapless corner mode, corresponding to $\Gamma_2$ in the same phase in which $\Gamma_1$ is gapless.
	
	The gapped $\Gamma_3$ corner can also be accounted in the similar way. From the relation $\mathcal{C}_3=\tilde{\mathcal{C}}_6^2$, by gluing the two cylinders in Fig. 5, we can obtain the compositional ``pumping cylinder" for $\mathcal{C}_3$, shown in Fig. 6. The result shows the two eigenvalues return to their original values after the pumping, which is not surprising since $\mathcal{C}_3$ is unitary. One of them circles around the cylinder, while the other one moves trivially. The net crossing number can be read out  from Fig. 6 to be one, which, according to our principle, should produce a gapless corner state. However, we should remember that from the result of the bands in ribbon geometry, there is virtually a real band crossing happening around the corner $\Gamma_3$. In view that each band of the Wannier Hamiltonian corresponds to each boundary modes in the ribbon bands, the phase produced in the ribbon band crossing compensates the phase produced by the circling of one of the eigenvalue, which is quite similar to the Laughlin's argument\cite{laughlin1999nobel,avron2003a} for quantized Hall conductances in quantum Hall systems. So the angle $\Gamma_3$ is indeed gapped.
	
	\begin{figure}
	\centering
	\includegraphics[width=8cm]{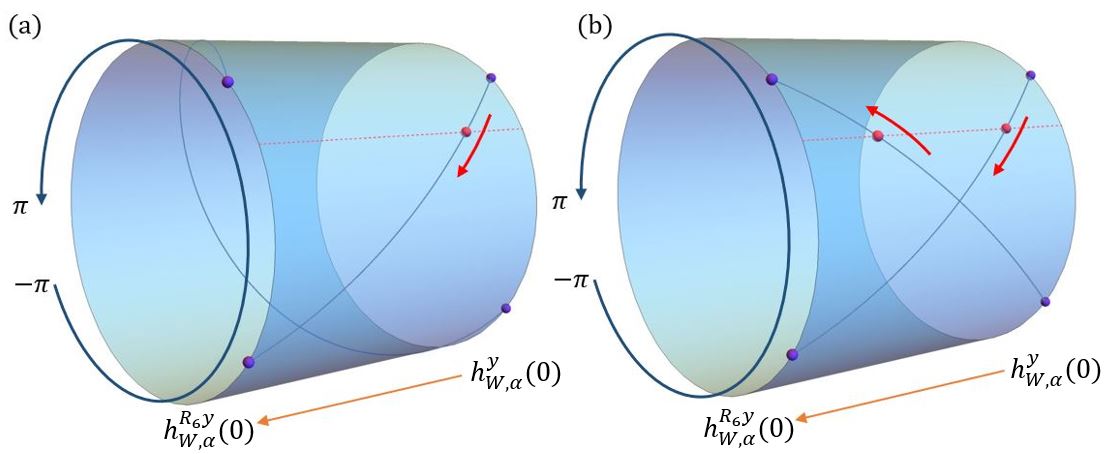}
	\caption{The two possible situations of band flip in the Wannier Hamiltonian at point $k_x=0$. (a)This crossing way has exactly one crossing at the reference line, thus producing the Chiral corner modes similar to the 1D boundary of quantum Hall systems. (b) No matter what reference point we choose, this situation have even number of total crossings, thus the corner produced by rotation in position space does not host any gapless corner modes.}
	\end{figure}
	
	It is worth pointing out that symmetry transformation between the two edges of the corner and the ``pumping cylinder", which sews the Wannier Hamiltonian before and after the transformation, does not have a one-to-one correspondence. This is readily seen from difference between the $\tilde{\mathcal{C}}_6^*$ rotation at the $\Gamma_2$ corner from $-\pi/3$ to $0$ (we use a star to indicate this transformation only transforms between these two angles) and the the rotation of $\Gamma_1$ from 0 to $\pi/3$. Similar things happen to $\mathcal{C}_3$, because the transformation $\mathcal{C}_3^A$ from $0$ to the angle $2\pi/3$ starts from a bond with hopping $t_2$ and ends at $t_1$, but the transformation  $\mathcal{C}_3^B$ from $-\pi/3$ to $\pi/3$ starts and ends conversely. In Appendix B we shall see they indeed lead to mutually inverse cylinder correspondence. Thus, to avoid ambiguity, when we refer to a transformation here, we should also specify it's action in position space. 

	\begin{figure}
	\centering
	\includegraphics[width=6.5cm]{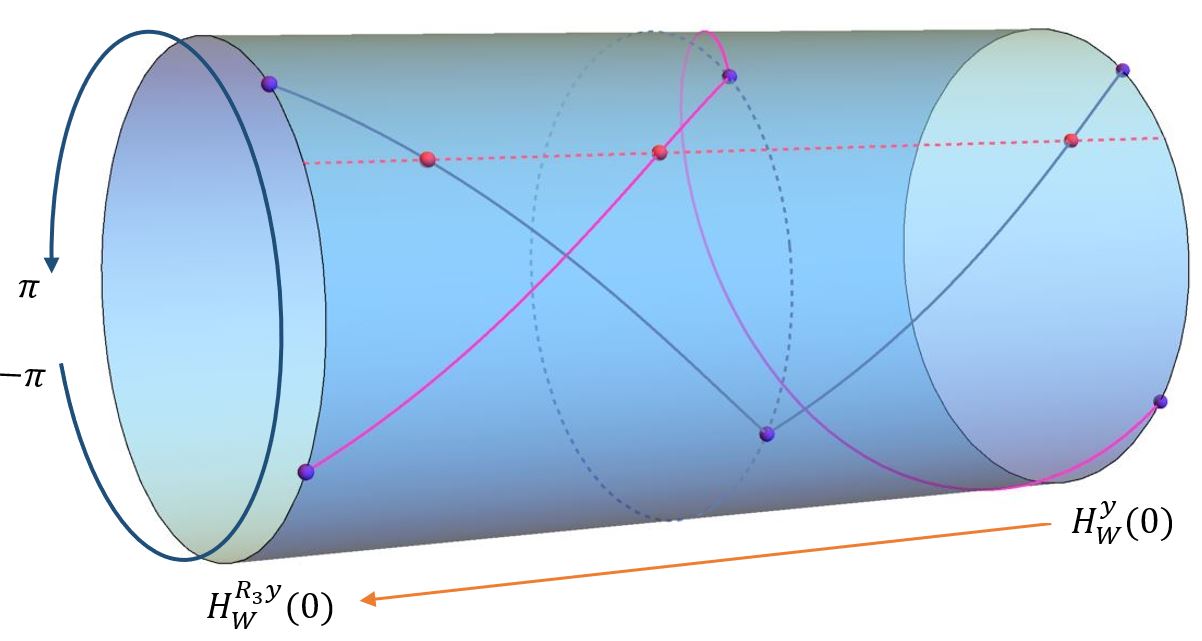}
	\caption{The concatenated pumping cylinder between the two edges of the $\Gamma_3$ corner. The route indicated by the purple line goes around the cylinder for one time, producing a flux for the eigenvalue pump in the bands of the ribbon geometry.}
\end{figure}
	\section{conclusion}
	The chiral HOTI on breathing Kagome lattice has revealed crucial distinctions between different inequivalent corners on the lattice, which former publications, using square lattice, do not incorporate. This dependence originates from the breaking of rotational symmetries, which can only be recovered to become anti-unitary. We have shown a generalized relation that if certain rotation symmetry through the corner on 2D lattice systems has been explicitly broken and can only be restored by a compositional anti-unitary symmetry, the corner corresponding to this rotation symmetry can conditionally host gapless corner states. Our proof, using the Wilson loop formalism, captures the ``band inversion" happening in the eigenspace of the Wannier Hamiltonian. This symmetry/gapless correspondence is not only true for the model in our paper, in which the anti-unitary nature of $\tilde{\mathcal{C}}_6$ corresponds to the gapless-ability of $\Gamma_{1,2}$, but rather a general scheme in determining the existence of chiral gapless corner modes in 2D and hinge modes in higher dimensions. We believe this methods is capable in exploring the existence of chiral majorana zero modes in the higher order topological superconductors, and to guide our prediction of realistic models with higher topological properties. 
	
	It's enlightening to compare our arguments using Wilson loop and the classification of Floquet topological phases, in both of which the Hamiltonians (which is the Wannier Hamiltonian for HOTI and the real Hamiltonian for Floquet systems) have certain periodicities\cite{von2016phase,potter2016classification} in their target space. This similarity could possibly be used to establish a correspondence between a $d$ dimensional second-order topological phase and a $d-1$ dimensional Floquet system. This potential correspondence is promising under future investigation.
	
	\begin{acknowledgments}
		YX would like to thank Fei Song, Zhong Wang and Zhongbo Yan for useful discussions, especially in bringing the concept of nested Wilson loops to the work. YX and RX are supported by Yan Jici Sci-Tech Elite Class of Physics, USTC. SW is supported by NSFC under Grant No.11275180.
	\end{acknowledgments}

	\appendix
	
	\section{Classification of the corner states through the ``pumping cylinder"}
	Here we demonstrate that there is indeed two possibilities of the ``pumping cylinder" defined in the main text in Fig. 5, thus having a two-fold classification. To do so, we first give a similar proof for the classification to the case with a unitary symmetry operation, thus the eigenvalues are not flipped when we go through the cylinder. The proof bases on the two essential observations below (in the discussions below we omit the axis $[-\pi,\pi)$ and the arrow, which we implicitly mean the pumping is from the right to the left of the cylinder):
	\begin{enumerate}
		\item If both the pumping of the two eigenvalues circle the cylinder once, the total result is trivial. (See Fig. 7.)
		\begin{figure}[H]
			\centering
			\includegraphics[width=8.5cm]{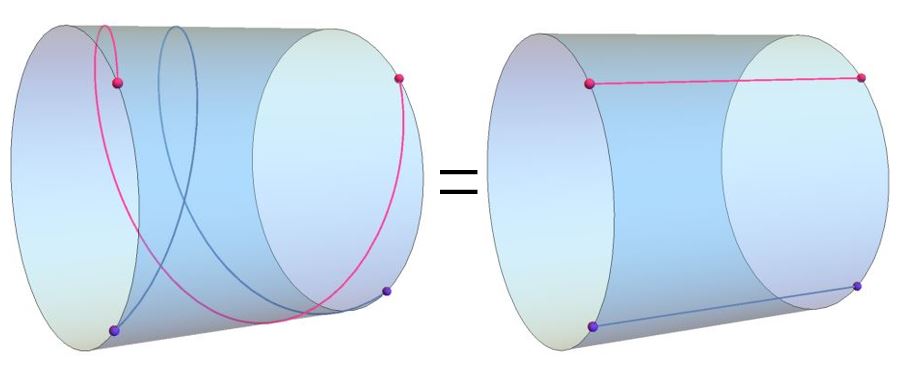}
			\caption{The equivalence of a total circling and trivial pumping}
		\end{figure}
		
		This can be easily seen from the triviality of adding a total phase, which, in our case, is trivial, because on the cylinder only the difference between the two eigenstates is meaningful. This means in the discussion below we can set one of the eigenvalue to be fixed during the entire pump to simplify our results. 
		\item If an eigenvalue circles the cylinder for three times during the pump, the total result is trivial.
			\begin{figure}[H]
			\centering
			\includegraphics[width=8.5cm]{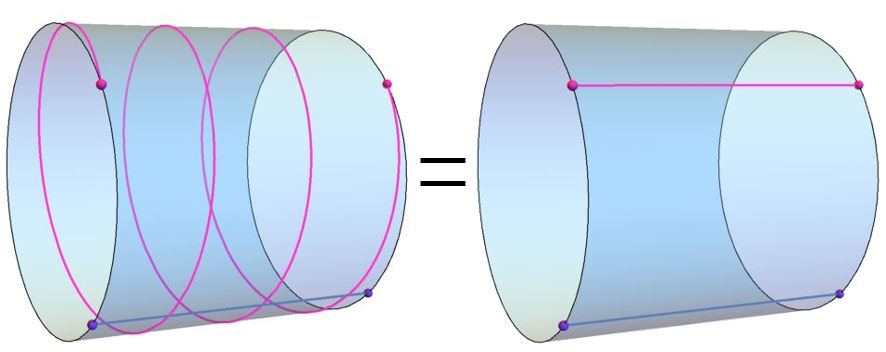}
			\caption{The equivalence of a triple circling and trivial pumping}
		\end{figure}
		This feature originates from the identity $\mathcal{C}_3^3=1$, together with the unitary nature of $\mathcal{C}_3$, that the only possibility of the pumping cylinder is one of the eigenvalues circling the cylinder for $n\in \mathbb{Z}$ times (and the other one fixed, from the discussion in 1.). For each choice, there must be a equivalence relation $3n\sim 0$ from the identity, thus we must have $3\sim 0$, which is just the equivalence in Fig. 8.
	\end{enumerate}
	Now we can apply these two results into the classification of a pumping cylinder with eigenvalue crossings. We denote the anticlockwise circling of the upper eigenvalue around the cylinder once/twice as $S_1/S_2$, and the no circling case $I$. Then $\{I,S_1,S_2\}$ form a $C_3$ group. $S_1$, the generator of the group, can be used now to generate the pumping cylinder in the case with band inversion. We choose the pumping of the trivial process in Fig. 5(b) as the base point to be generated. If we denote it as $X$, then clearly Fig. 5(a) is just the first in the generation $S_1X$. The second generation $S_2X$, visualized in Fig. 9, is equivalent a crossing on the other side of the cylinder.
	\begin{figure}[H]
		\centering
		\includegraphics[width=8.5cm]{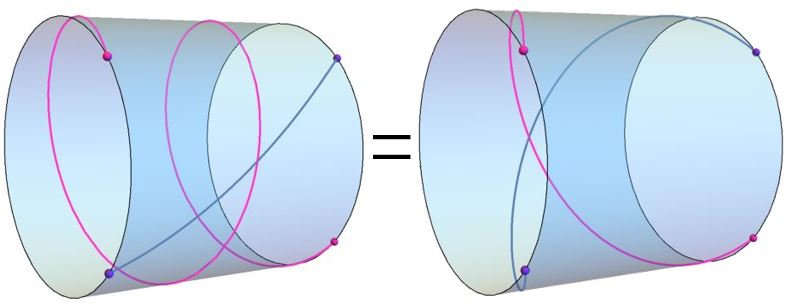}
		\caption{The generation of $S_2X$ and the equivalence to a trivial cross.}
	\end{figure}
	This equivalence is from the result 1 in the discussion above. The result give us an interesting conclusion that although we have the operator group $C_3$ generated by $S_1$, the number of different pumping cylinders in the band crossing case is two. This proves our claim in the main text.
	\section{The correspondence between the pumping cylinder and corners}
	It's intriguing to visualize physically our pumping cylinders, i.e. to find representatives of these pumping cylinders in different inequivalent corners in our Kagome lattice model. We have already known   from our numerical results in the main text that in the phase with gapless $\Gamma_1$ ($-t_2<t_1<t_2/2$), the angle $\Gamma_1$ and $\Gamma_2$ corresponds to $S_1X$ and $X$, while $\Gamma_3$ the rotation $\mathcal{C}_3^A$ from 0 to $\pi/3$ corresponds to $S_2$ (see Fig. 6 and result 2 in Appendix A). The representative of the rotation $\mathcal{C}_3^B$ is not equivalent to $S_2$, but rather $S_1=S_2^{-1}$, because the pumping process is reversed, i.e. from a corner with intersecting hoppings $t_2\to t_1$ to $t_1\to t_2$. This observation is also consistent with the generation of inequivalent crossing diagrams in Appendix A, in which we compose the representatives of $\tilde{\mathcal{C}}_6^{-1}$ with $\mathcal{C}_3^B$, $S_1$ and $X$, to get the pumping cylinder for $\tilde{\mathcal{C}}_6$, $S_1X$. 
	\begin{figure}[H]
		\centering
		\includegraphics[width=5.5cm]{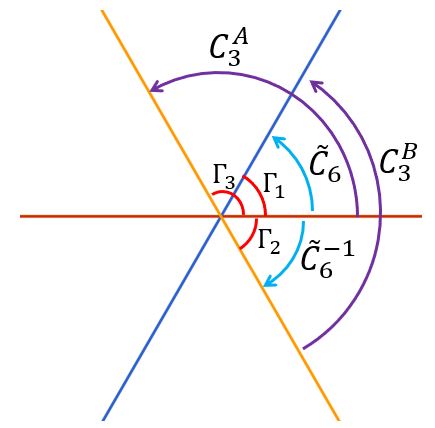}
		\caption{The different symmetry operations and corners.}
	\end{figure}
	In the phase with gapless $\Gamma_2$ ($-t_1<t_2<t_1/2$), similar reasoning will show $\Gamma_1$ and $\Gamma_2$ are now correspond to $X$ and $S_1X$, with $C_3^A$ corresponds to $S_1$. In the trivial gapped phase $t_1+t_2<0$ all the $\pi/6$ corners corresponds to $X$ and all the $C_3$ operations corresponds to $I$ in the cyclic group. Thus, in the three gapped phase in the phase diagram Fig. 2(c) the operation $\mathcal{C}_3^A$ is ``represented" by three different elements in the cylinder cyclic group ${I,S_1,S_2}$. This interesting feature shows while the bulk of the phase is $\mathbb{Z}_3$ classified, each angle state possesses two-fold classification.

	\bibliography{bib}
	
\end{document}